# Nucleation kinetics of vapor bubbles in a liquid with arbitrary viscosity

Nikolay V. Alekseechkin

Akhiezer Institute for Theoretical Physics, National Science Centre "Kharkov Institute of Physics and Technology", Akademicheskaya Street 1, Kharkov 61108, Ukraine

**Abstract.** The theory of vapor bubbles nucleation in single-component liquids developed in [J. Phys. Chem. B **116**, 9445 (2012)] for the case of high viscosity (the ($V, \rho, T$)- theory) is extended to the case of arbitrary viscosity. For this purpose, Langevin's approach in the theory of Brownian motion, or Kramers' model of chemical reactions, is employed. The obtained expression for the bubbles nucleation rate is valid for arbitrary relations between the kinetic parameters controlling the nucleation process: viscosity, inertia of a liquid, the rate of evaporation into a bubble and the rate of heat exchange between the bubble and ambient liquid. So, the presented theory together with the ($V, \rho, T$)- theory gives a complete description of the vapor-bubbles nucleation kinetics in one-component liquids. Limiting cases with respect to the mentioned parameters are considered, in particular, the low viscosity limit. It is shown that the low- and high-viscosity nucleation rates differ from each other qualitatively and quantitatively. The possibility of application of the theory to cavitation in superfluid helium-4 is discussed.

## 1. Introduction

The kinetics of bubbles nucleation in metastable liquids [1-11] is intensively studied both theoretically and experimentally. The range of experimental conditions for bubbles formation is very wide; in particular, it extends with respect to temperature to values of the order of $1\,°K$ for superfluid helium-4, where the thermal nucleation of bubbles is also investigated [11-17]. The theoretical basis for studying this phenomenon is the classical theory of nucleation [1, 2, 18-20] generalized to the case of several variables describing the embryo state [21-25] and improved in various aspects [26-34]. In particular, the studies of the pre-exponential factor in the equilibrium distribution function of nuclei carried out in Refs. [27-30] are very important for a correct calculation of the nucleation rate. The density functional approach [31-34] as an alternative to the Gibbs capillarity theory has been developed for calculating the critical-nucleus free energy and thereby the nucleation rate in the cases of small critical sizes, where the application of the capillarity approach requires the knowledge of some important dependences such as the dependences of the surface tension and intensive state parameters of a nucleus on its radius [35].

It was shown recently [35] that a multivariable theory coupled with macroscopic kinetics provides the most complete and comprehensive description of the kinetics of vapor bubbles nucleation in a metastable liquid. This description takes into account both thermodynamic and kinetic properties of a liquid and allows to consider limiting effects of the latter on the nucleation process. The Gibbs capillarity theory employed in Ref. [35] is natural for the multivariable treatment of the problem: the work of nucleus formation naturally appears therein as a function of all variables describing the bubble state.

The theory of Ref. [35] relates to the case of high viscosity $\eta_0$ of a liquid (or high "friction coefficient" [25] $\gamma_0 \sim \eta_0$). Note that the same is true for the classical theory of Zeldovich and Frenkel [2, 20]; Zeldovich notes this fact in Ref. [2], so the Smoluchowski equation of the Brownian motion theory is obtained for the distribution function of clusters [2, 20]. The Einstein-Smoluchowski theory of Brownian motion as a succinct description of this process is the limiting case of a more general Langevin's approach [36, 37], where the case of arbitrary damping (or arbitrary viscosity) is considered. Einstein's description of Brownian motion uses a single variable – the position $\mathbf{r}$ of a particle. Langevin's approach deals with mechanical equations of motion of a Brownian particle of mass $M$, accordingly, the velocity $\mathbf{u} = \dot{\mathbf{r}}$ is also a variable of the theory; ($\mathbf{r}, \mathbf{u}$) is the set of variables of the Fokker-Planck equation [36, 37].

Just Langevin's approach is employed by Kramers [19] to consider the transition of a particle over a potential barrier. Kramers' model is also a model of nucleation, therefore the limiting cases of high and low viscosity considered therein also have to occur in nucleation Phenomena [7, 25].

In terms of the multivariable nucleation theory [24, 25], the velocities $u_i$ are stable variables, since their contribution to the work of embryo formation is the positive "kinetic energy" $m_{ik} u_i u_k / 2$, where $m_{ik}$ is the "mass tensor" [25]. In Ref. [25], the lifetime of a Brownian particle in a potential well is calculated based on the formula for the steady-state nucleation rate in a multivariable nucleation theory. However, the following distinction of a usual Brownian particle from an embryo must be noted. Coordinates $x_i$ of the Brownian particle are physically equivalent (though one of them is unstable for the process of escape of the particle from a potential well, as in the nucleation theory); the quantities $x_i$ may be the usual Cartesian coordinates or some generalized ones. Accordingly, the number of velocities $u_i$ is equal to the number of coordinates $x_i$. In the nucleation theory which is in essence the theory of processes with linked fluxes [24, 38, 39] (as shown in Ref. [24], the processes of binary and multicomponent nucleation also can be treated as linked-fluxes processes), the variables describing an embryo have different physical meaning. In our case, these are the bubble volume $V$ (an unstable variable), the vapor density $\rho$ and temperature $T$ (stable variables). The changes of stable variables, $\delta\rho$ and $\delta T$, contain a part proportional to the change $\delta V$ of the unstable variable [24, 35], according to the linked-fluxes analysis. Thus, only one "velocity" $u = \dot{V}$ will replenish the set of stable variables in Langevin's treatment of the process of bubbles formation.

The method of calculating the steady-state nucleation rate in Ref. [24] does not require the symmetry of the kinetic matrix **D** (the matrix of "diffusivities" in the Fokker-Planck equation); this matrix contains antisymmetric elements, $d_{x_i u} = -d_{ux_i}$, $x_i = \{V, \rho, T\}$, in Langevin's approach. Accordingly, the nucleation-rate formula obtained therein and used in Ref. [35] is valid in Langevin's approach also.

The paper is organized as follows. The equations of motion of a bubble in the phase space $\{V, \rho, T, u\}$ are written in Sec. 2 and the kinetic matrix **D** is derived. Further, the symmetry conditions are applied to the matrix **D** and the matrix **Z** of the equations of motion is obtained in explicit form. The characteristic equation of the matrix **Z** and its solution are presented in Sec. 3. Various limiting cases of this solution are considered in Sec. 4 and Appendices. They correspond to specific relations between the kinetic parameters determining the nucleation rate: viscosity, inertia of the liquid, the rate of evaporation into the bubble and the rate of heat exchange between the bubble and the liquid. The summary of results is given in Sec. 5.

## 2. Equations of motion of vapor bubble

### 2.1. Equations of motion in general form and conditions of symmetry of kinetic coefficients

Thermodynamics of a nucleation process has been considered in Refs. [35, 40]. The work of embryo formation has the form $W = W_* + (1/2)\sum_{i,k} h_{ik}(x_i - x_i^*)(x_k - x_k^*)$ near the saddle point; the matrix **H** has been calculated in Ref. [35] in the approximation of perfect gas for the vapor. It should be noted that the diagonal form of the matrix **H** in the variables $(V, \rho, T)$ does not depend on this approximation. Suppose we have an arbitrary (real) gas with the equation of state

$$F(P,\rho,T) = 0 \tag{1}$$

The equation for the second differential of $W$ obtained in Ref. [35],

$$(d^2W)_* = N_*[dTds - dPd\upsilon]_* - \frac{2}{9}g\sigma V_*^{-4/3}(dV)^2 \tag{2}$$

is necessary to calculate the matrix $\mathbf{H}$. Here $s$ and $\upsilon$ are the entropy and the volume per one molecule in the bulk vapor phase, $N$ is the number of molecules in the bubble, $\sigma$ is the surface tension and $g = 3^{2/3}(4\pi)^{1/3}$; asterisk denotes the saddle-point value. The quantities $s$ and $P$ in Eq. (2) are functions of $\rho$ and $T$. So,

$$dP = \left(\frac{\partial P}{\partial \rho}\right)_T d\rho + \left(\frac{\partial P}{\partial T}\right)_\rho dT \tag{3}$$

The partial derivatives in this equation are determined from Eq. (1). In the similar equation for $ds$, we employ the following thermodynamic relations:

$$\left(\frac{\partial s}{\partial P}\right)_T = -\left(\frac{\partial \upsilon}{\partial T}\right)_P, \quad \left(\frac{\partial s}{\partial \rho}\right)_T = \left(\frac{\partial s}{\partial P}\frac{\partial P}{\partial \rho}\right)_T = -\left(\frac{\partial \upsilon}{\partial T}\right)_P\left(\frac{\partial P}{\partial \rho}\right)_T, \quad \left(\frac{\partial s}{\partial T}\right)_\rho = \left(\frac{\partial s}{\partial T}\right)_V = \frac{c_V}{T},$$

$$\frac{\partial \upsilon}{\partial T} = -\frac{1}{\rho^2}\frac{\partial \rho}{\partial T} \tag{4}$$

where $c_V$ is the vapor heat capacity per one molecule.

Substituting Eq. (3) and the similar equation for $ds$ with Eqs. (4) in Eq. (2), we find the term at $d\rho dT$, i.e. the element $h_{\rho T}$:

$$h_{\rho T} = \frac{N_*}{\rho^2}\left\{\left(\frac{\partial P}{\partial T}\right)_\rho + \left(\frac{\partial \rho}{\partial T}\right)_P\left(\frac{\partial P}{\partial \rho}\right)_T\right\} = 0 \tag{5}$$

in view of the known mathematical identity for Eq. (1)

$$\left(\frac{\partial P}{\partial \rho}\right)_T\left(\frac{\partial \rho}{\partial T}\right)_P\left(\frac{\partial T}{\partial P}\right)_\rho = -1 \tag{6}$$

So, the matrix $\mathbf{H}$ has a canonical form

$$\mathbf{H} = \begin{pmatrix} -\frac{2}{9}g\sigma V_*^{-\frac{4}{3}} & 0 & 0 \\ 0 & \frac{(\partial P/\partial \rho)_T^* V_*}{\rho_*} & 0 \\ 0 & 0 & \frac{c_V \rho_* V_*}{T_0} \end{pmatrix} \tag{7}$$

for arbitrary gas also.

As for the elements of the matrix $\mathbf{Z}$ [35] (it is considered below), the element $z_{VV}$ does not change, the element $z_{V\rho}$ is equal now to $-3\xi V_*(\partial P/\partial \rho)_T^*$. Hence, the coefficient $a_\rho = z_{V\rho}h_{VV}/z_{VV}h_{\rho\rho} = -\rho_*/V_*$, as before.

Nevertheless, the approximation of perfect gas is employed in the present report for simplicity and clarity of the theory. Recalculation of the nucleation rate for the case of a real gas can be performed directly following the algorithm of the work. The critical radius $R_*$ is determined from the condition of mechanical equilibrium

$$P_*(R_*) - P_0 = \frac{2\sigma(R_*)}{R_*} \tag{8}$$

where the dependence $P_*(R_*)$ in the cases of a real gas and sufficiently small values of $R_*$ is not given by the Kelvin formula [35]; it can be found from general equations of Ref. [41]. Here $P_0$ and $T_0$ are the pressure and the temperature in the liquid.

Returning to the work $W$, as noted above, it is complemented by the "kinetic energy" $Mu^2/2$ in Langevin's approach; accordingly, the matrix $\mathbf{H}$ of Ref. [35] is complemented by the "mass" $M$ of the bubble:

$$\mathbf{H} = \begin{pmatrix} -\frac{2}{9}g\sigma V_*^{-\frac{4}{3}} & 0 & 0 & 0 \\ 0 & \frac{kT_0 V_*}{\rho_*} & 0 & 0 \\ 0 & 0 & \frac{c_V \rho_* V_*}{T_0} & 0 \\ 0 & 0 & 0 & M \end{pmatrix} \tag{9}$$

Equations of motion of a bubble [35] in the $(V,\rho,T,u)$-theory have the following general form:

$$\begin{cases} \dot{V} = u = -z_{Vu}u \\ \dot{\rho} = -z_{\rho V}(V-V_*) - z_{\rho\rho}(\rho-\rho_*) - z_{\rho T}(T-T_0) - z_{\rho u}u \\ \dot{T} = -z_{TV}(V-V_*) - z_{T\rho}(\rho-\rho_*) - z_{TT}(T-T_0) - z_{Tu}u \\ \dot{u} = -z_{uV}(V-V_*) - z_{u\rho}(\rho-\rho_*) - z_{uT}(T-T_0) - z_{uu}u \end{cases} \tag{10}$$

From here, the matrix $\mathbf{D} = kT_0 \mathbf{Z}\mathbf{H}^{-1}$ of diffusivities in the Fokker-Planck equation [24] is calculated:

$$\frac{\mathbf{D}}{kT_0} = \begin{pmatrix} 0 & 0 & 0 & -M^{-1} \\ z_{\rho V}h_{VV}^{-1} & z_{\rho\rho}h_{\rho\rho}^{-1} & z_{\rho T}h_{TT}^{-1} & z_{\rho u}M^{-1} \\ z_{TV}h_{VV}^{-1} & z_{T\rho}h_{\rho\rho}^{-1} & z_{TT}h_{TT}^{-1} & z_{Tu}M^{-1} \\ z_{uV}h_{VV}^{-1} & z_{u\rho}h_{\rho\rho}^{-1} & z_{uT}h_{TT}^{-1} & z_{uu}M^{-1} \end{pmatrix} \tag{11}$$

The conditions of symmetry of the matrix **D** read as follows:

$$z_{\rho V} = 0, \quad z_{TV} = 0 \tag{12a}$$

$$z_{T\rho} h_{\rho\rho}^{-1} = z_{\rho T} h_{TT}^{-1} \tag{12b}$$

for symmetric elements and

$$z_{uV} h_{VV}^{-1} = M^{-1} \tag{13a}$$

$$z_{u\rho} h_{\rho\rho}^{-1} = -z_{\rho u} M^{-1} \tag{13b}$$

$$z_{uT} h_{TT}^{-1} = -z_{Tu} M^{-1} \tag{13c}$$

for antisymmetric elements.

## 2.2. Deriving the matrix Z

Denoting $z_{\rho u} \equiv -a_\rho$, $z_{Tu} \equiv -a_T$, $z_{\rho\rho} \equiv \lambda_{\rho\rho}$, $z_{\rho T} \equiv \lambda_{\rho T}$, $z_{T\rho} \equiv \lambda_{T\rho}$, and $z_{TT} \equiv \lambda_{TT}$, we see that Eqs. (10) for $\dot\rho$ and $\dot T$, in view of Eq. (12a), acquire the same form,

$$\dot\rho = a_\rho \dot V - \lambda_{\rho\rho}(\rho - \rho_*) - \lambda_{\rho T}(T - T_0) \tag{14a}$$

$$\dot T = a_T \dot V - \lambda_{T\rho}(\rho - \rho_*) - \lambda_{TT}(T - T_0) \tag{14b}$$

as in Ref. [35].

Eqs. (13a)-(13c) determine the mass $M$ and the coefficients $a_\rho$ and $a_T$ by the known equation for $\dot u$. This equation is given in Ref. [35]:

$$\dot u = \ddot V = \frac{\dot V^2}{6V} - \frac{4\eta_0}{b\rho_0} \frac{\dot V}{V^{2/3}} + \frac{3}{b\rho_0} V^{1/3}\left(P - \frac{2\sigma}{\sqrt{b}V^{1/3}} - P_0\right) \tag{15}$$

where $\rho_0$ is the mass density of a liquid, $b = (3/4\pi)^{2/3}$, and

$$P = \rho k T \tag{16}$$

is the vapor pressure in the bubble.

We neglect by the term $\sim \dot V^2$ in Eq. (15) and expand the right side up to linear terms near the saddle point:

$$\dot u = \frac{8\pi}{3}\frac{\sigma}{\rho_0 V_*}(V - V_*) + \frac{3kT_0 V_*^{1/3}}{b\rho_0}(\rho - \rho_*) + \frac{3k\rho_* V_*^{1/3}}{b\rho_0}(T - T_0) - \frac{4\eta_0 V_*^{-2/3}}{b\rho_0} u \tag{17}$$

Thus,

$$z_{uV} = -\frac{8\pi}{3}\frac{\sigma}{\rho_0 V_*}, \quad z_{u\rho} = -\frac{3kT_0 V_*^{1/3}}{b\rho_0}, \quad z_{uT} = -\frac{3k\rho_* V_*^{1/3}}{b\rho_0}, \quad z_{uu} = \frac{4\eta_0 V_*^{-2/3}}{b\rho_0} = \frac{1}{\xi\rho_0 R_*^2}, \quad \xi \equiv \frac{1}{4\eta_0} \tag{18}$$

Eq. (13a) gives for the mass $M$ the following value:

$$M = \frac{h_{VV}}{z_{uV}} = \frac{b}{3}\rho_0 V_*^{-1/3} = \frac{\rho_0}{4\pi R_*} \tag{19}$$

The coefficients $a_\rho$ and $a_T$, according to Eqs. (13b) and (13c), are

$$a_\rho = \frac{z_{u\rho}M}{h_{\rho\rho}} = -\frac{\rho_*}{V_*}, \quad a_T = \frac{z_{uT}M}{h_{TT}} = -\frac{P_*}{C_V^*} \tag{20}$$

where $C_V^* = c_V \rho_* V_*$ is the heat capacity of the critical bubble.

So, the same values of the coefficients $a_\rho$ and $a_T$ as in Ref. [35] are obtained. Equations for $\dot\rho$ and $\dot T$ in the form of Eqs. (14a) and (14b) have been written in Ref. [35] from intuitive considerations based on the linked-fluxes concept [38, 39]. We see that the more general $(V,\rho,T,u)$-theory gives just such a representation of the equations for $\dot\rho$, $\dot T$ and determines the coefficients $a_\rho$ and $a_T$. This fact shows the self-consistency of the theory. The coefficients $\lambda_{ik}$ are calculated in Ref. [35]; Eq. (12b) is Onsager's reciprocal relation for them. The above consideration also shows the connection between the symmetry properties of the matrix **D** and the linked-fluxes analysis.

Now the matrix **Z** is completely determined:

$$\mathbf{Z} = \begin{pmatrix} 0 & 0 & 0 & -1 \\ 0 & \lambda_{\rho\rho} & \lambda_{\rho T} & -a_\rho \\ 0 & \lambda_{T\rho} & \lambda_{TT} & -a_T \\ z_{uV} & z_{u\rho} & z_{uT} & z_{uu} \end{pmatrix} \tag{21}$$

$$\det \mathbf{Z} = z_{uV} \det \mathbf{\Lambda} \tag{22}$$

## 3. Characteristic equation and its solution

The characteristic equation for the matrix **Z** has the following form:

$$\kappa^4 - (Sp\mathbf{Z})\kappa^3 + B_2\kappa^2 - B_3\kappa + \det \mathbf{Z} = 0 \tag{23}$$

The coefficients in this equation are the invariants of the matrix **Z**: $Sp\mathbf{Z}$ is the trace, $B_2$ and $B_3$ are the sums of the principal minors of the second and third orders, respectively, and the determinant. In view of the following relations

$$z_{uV} = (-\xi P_L^*)z_{uu} = z_{VV} z_{uu} \tag{24a}$$

$$a_T z_{uT} = \left(\frac{3k}{c_V}\xi P_*\right) z_{uu} \tag{24b}$$

$$a_\rho z_{u\rho} = (3\xi P_*)z_{uu} \tag{24c}$$

$$a_T z_{u\rho} \lambda_{\rho T} = a_\rho z_{uT} \lambda_{T\rho} = -\left(3\xi \frac{k}{c_V} P_*\right) \tilde{q} \lambda_{\rho\rho} \tag{24d}$$

where $P_L^* = 2\sigma/R_*$, $\tilde{q} = [q - kT_0 + 2\upsilon_0 \sigma/R_*]/kT_0$, $q$ is the heat of evaporation per one molecule, $\upsilon_0$ is the volume per one molecule in the liquid, the coefficients of Eq. (23) are represented as

$$Sp\mathbf{Z} = \lambda_{\rho\rho} + \lambda_{TT} + z_{uu} \tag{25a}$$

$$B_2 = z_{uu} Sp\mathbf{Z}_{(3)} + \det \mathbf{\Lambda} \tag{25b}$$

$$B_3 = Bz_{uu} \tag{25c}$$

$$\det \mathbf{Z} = z_{uu} \det \mathbf{Z}_{(3)} \tag{25d}$$

where $\mathbf{Z}_{(3)}$ is the matrix $\mathbf{Z}$ of the three-variable $(V, \rho, T)$-theory [35],

$$\mathbf{Z}_{(3)} = \begin{pmatrix} -\xi P_L^* & -3\xi V_* kT_0 & -3\xi kN_* \\ \xi P_L^* \dfrac{\rho_*}{V_*} & \lambda_{\rho\rho} + 3\xi P_* & \lambda_{\rho T} + 3\xi k\rho_*^2 \\ \xi P_L^* \dfrac{P_*}{C_V^*} & \lambda_{T\rho} + 3\xi \dfrac{(kT_0)^2}{c_V} & \lambda_{TT} + 3\xi \dfrac{k}{c_V} P_* \end{pmatrix} \tag{26}$$

$$Sp\mathbf{Z}_{(3)} = -\xi P_L^* + \lambda_{\rho\rho} + \lambda_{TT} + 3\left(1 + \frac{k}{c_V}\right)\xi P_* \tag{27a}$$

$$\det \mathbf{Z}_{(3)} = z_{VV} \det \mathbf{\Lambda} \tag{27b}$$

$$B = z_{VV}(\lambda_{\rho\rho} + \lambda_{TT}) + \det \mathbf{\Lambda} + (1 + 2\tilde{q})\left(3\xi \frac{k}{c_V} P_*\right)\lambda_{\rho\rho} + 3\xi P_* \lambda_{TT} \tag{27c}$$

So, the coefficients of Eq. (23) are expressed via the coefficients $Sp\mathbf{Z}_{(3)}$, $B$ and $\det \mathbf{Z}_{(3)}$ of the characteristic equation of the $(V, \rho, T)$-theory

$$\kappa^3 - (Sp\mathbf{Z}_{(3)})\kappa^2 + B\kappa - \det \mathbf{Z}_{(3)} = 0 \tag{28}$$

We have to select the negative root $\kappa_1$ from four roots of Eq. (23). In the Descartes-Euler solution [42], the desired root has the following form:

$$\kappa_1 = -\sqrt{y_1} - \sqrt{y_2} + \sqrt{y_3} + \frac{Sp\mathbf{Z}}{4} \tag{29}$$

where

$$y_1 = 2\sqrt{-\frac{p'}{3}} \cos\frac{\phi'}{3} - \frac{a'}{3} \tag{30a}$$

$$y_2 = -2\sqrt{-\frac{p'}{3}} \cos\left(\frac{\phi'}{3} + \frac{\pi}{3}\right) - \frac{a'}{3} \tag{30b}$$

$$y_3 = -2\sqrt{-\frac{p'}{3}}\cos\left(\frac{\phi'}{3} - \frac{\pi}{3}\right) - \frac{a'}{3} \tag{30c}$$

with

$$p' = b' - \frac{a'^2}{3}, \quad q' = \frac{2}{27}a'^3 - \frac{1}{3}a'b' + c', \quad \cos\phi' = -\frac{q'}{2\sqrt{-(p'/3)^3}} \tag{31}$$

are the solutions of the cubic equation

$$y^3 + a'y^2 + b'y + c' = 0 \tag{32}$$

with

$$a' = \frac{p}{2}, \quad b' = \frac{p^2 - 4r}{16}, \quad c' = -\frac{q^2}{64} \tag{33}$$

and

$$p = -\frac{3}{8}(Sp\mathbf{Z})^2 + B_2 \tag{34a}$$

$$q = -\frac{1}{8}(Sp\mathbf{Z})^3 + \frac{1}{2}(Sp\mathbf{Z})B_2 - B_3 \tag{34b}$$

$$r = -\frac{3}{256}(Sp\mathbf{Z})^4 + \frac{1}{16}(Sp\mathbf{Z})^2 B_2 - \frac{1}{4}(Sp\mathbf{Z})B_3 + \det\mathbf{Z} \tag{34c}$$

From Eqs. (31) and (33),

$$p' = -\frac{1}{4}\left(\frac{p^2}{12} + r\right) \tag{35a}$$

$$q' = -\frac{1}{8}\left(\frac{p^3}{108} - \frac{pr}{3} + \frac{q^2}{8}\right) \tag{35b}$$

The validity of Eq. (29) is substantiated and confirmed by considering the limiting case of high viscosity in Appendix A.

The eigenvalue $\kappa_1$, Eq. (29), determines the steady-state nucleation rate of bubbles [35]

$$I = N_b \sqrt{\frac{kT_0}{2\pi}|h_{11}^{-1}|}\,|\kappa_1|\,e^{-\frac{W_*}{kT_0}} \tag{36}$$

where $N_b$ is the normalizing constant of the one-dimensional equilibrium distribution function $f_{eq}(V)$ of bubbles [27-30, 35].

## 4. Kinetic limits and discussion

As seen from the above, the presented $(V,\rho,T,u)$-theory contains the parameters $z_{VV}$, $\lambda_{\rho\rho}$, and $\lambda_{TT}$ of the $(V,\rho,T)$-theory and one more parameter $z_{uu}$.

### 4.1. Fast heat exchange limit

The limit of high viscosity, $z_{uu} \gg |z_{VV}|, \lambda_{\rho\rho}, \lambda_{TT}, 3\xi P_*$, is considered in Appendix A as a limiting case of Eq. (29). A much easier way that does not require the knowledge of the exact solution of Eq. (23) is employed here to deal with such limits. This method directly reduces the order of Eq. (23) in the case of a large value of any kinetic parameter. Such approach is especially useful in the case when the number of variables is greater than four, and there is no solution of the characteristic equation in radicals according to Abel's theorem.

Dividing Eq. (23) by $z_{uu}$ and considering the formal limit $z_{uu} \to \infty$ (the high viscosity limit), we see that a solution of the equation

$$\frac{1}{z_{uu}}\kappa^4 - \frac{Sp\mathbf{Z}}{z_{uu}}\kappa^3 + \frac{B_2}{z_{uu}}\kappa^2 - \frac{B_3}{z_{uu}}\kappa + \frac{\det \mathbf{Z}}{z_{uu}} = 0 \tag{37}$$

in this limit is also a solution of the equation

$$\frac{Sp\mathbf{Z}}{z_{uu}}\kappa^3 - \frac{B_2}{z_{uu}}\kappa^2 + \frac{B_3}{z_{uu}}\kappa - \frac{\det \mathbf{Z}}{z_{uu}} = 0 \tag{38}$$

From Eqs. (25a)-(25d), we find that in the limit $z_{uu} \to \infty$

$$\frac{Sp\mathbf{Z}}{z_{uu}} \to 1, \quad \frac{B_2}{z_{uu}} \to Sp\mathbf{Z}_{(3)}, \quad \frac{B_3}{z_{uu}} \to B, \quad \frac{\det \mathbf{Z}}{z_{uu}} \to \det \mathbf{Z}_{(3)} \tag{39}$$

so Eq. (38) is the characteristic equation of the $(V,\rho,T)$-theory, Eq. (28), as it must.

Let us consider similarly the limiting case of fast heat exchange $\lambda_{TT} \to \infty$ (in the three-variable $(V,\rho,T)$-theory, this limit corresponds to the transition to the two-variable $(V,\rho)$-theory [35]). Dividing Eq. (23) by $\lambda_{TT}$ and employing the relations

$$\frac{Sp\mathbf{Z}}{\lambda_{TT}} \to 1, \quad \frac{B_2}{\lambda_{TT}} \to z_{uu} + \lambda_{\rho\rho}, \quad \frac{B_3}{\lambda_{TT}} \to z_{uu}(z_{VV} + \lambda_{\rho\rho} + 3\xi P_*) = z_{uu}Sp\mathbf{Z}_{V\rho},$$

$$\frac{\det \mathbf{Z}}{\lambda_{TT}} \to z_{uu}z_{VV}\lambda_{\rho\rho} = z_{uu}\det \mathbf{Z}_{V\rho} \tag{40}$$

where

$$\mathbf{Z}_{V\rho} = \begin{pmatrix} z_{VV} & z_{V\rho} \\ a_\rho z_{VV} & \lambda_{\rho\rho} + a_\rho z_{V\rho} \end{pmatrix} \tag{41}$$

is the matrix of the $(V,\rho)$-theory, we get the following characteristic equation:

$$\kappa^3 - (z_{uu} + \lambda_{\rho\rho})\kappa^2 + (z_{uu} Sp\mathbf{Z}_{V\rho})\kappa - z_{uu} \det \mathbf{Z}_{V\rho} = 0 \tag{42}$$

The solution of this equation has the form similar to Eq. (A11) of Appendix A:

$$\kappa_1 = \frac{1}{3}\left\{-2\sqrt{(z_{uu} + \lambda_{\rho\rho})^2 - 3z_{uu} Sp\mathbf{Z}_{V\rho}} \cos\left(\frac{\phi}{3} - \frac{\pi}{3}\right) + (z_{uu} + \lambda_{\rho\rho})\right\} \tag{43}$$

where $\cos\phi$ is determined by Eq. (A7) with replacements $Sp\mathbf{Z}_{(3)} \to (z_{uu} + \lambda_{\rho\rho})$, $B \to z_{uu} Sp\mathbf{Z}_{V\rho}$, and $\det \mathbf{Z}_{(3)} \to z_{uu} \det \mathbf{Z}_{V\rho} = z_{uV}\lambda_{\rho\rho}$ in Eqs. (A5), (A6) for $Y$ and $U$.

### 4.2. Low viscosity and mass exchange parameter

Consider the asymptotic form of Eq. (43) in the case when both the parameters $z_{uu}$ and $\lambda_{\rho\rho}$ are small, $z_{uu} \sim \lambda_{\rho\rho} \ll |z_{VV}|$ and $|Sp\mathbf{Z}_{V\rho}| = |z_{VV} + 3\xi P_*| \sim |z_{VV}|$. In this case, $(z_{uu} + \lambda_{\rho\rho})^2 / z_{uu}|Sp\mathbf{Z}_{V\rho}| \sim z_{uu}/|z_{VV}| \ll 1$. Hence the solution given by Eq. (43) exists only if $Sp\mathbf{Z}_{V\rho} < 0$. This condition means that the pressure $P_0$ in the liquid has to be negative [35]:

$$P_0 < -\frac{2}{3}P_L^* \tag{44}$$

The asymptotic form of Eq. (43) for the given case can be written by analogy with Eq. (123) of Ref. [35], where the similar case of the $(V,\rho,T)$-theory was considered:

$$\kappa_1 = -\sqrt{z_{uu}|Sp\mathbf{Z}_{V\rho}|} \tag{45}$$

In the limit of high negative pressure, $|P_0| \gg P_*$, when $P_0 \approx -P_L^*$ and $Sp\mathbf{Z}_{V\rho} \to z_{VV}$, Eq. (45) becomes as follows:

$$\kappa_1 = -\sqrt{z_{uu}|z_{VV}|} = -\sqrt{|z_{uV}|} \tag{46}$$

i.e. the same equation is obtained, as in the case of small viscosity only, Eq. (B13) of Appendix B.

Thus, in the case of fast heat exchange, when both the parameters $z_{uu}$ and $\lambda_{\rho\rho}$ are small, the nucleation of bubbles is possible *under negative pressure* only; the pressure value has to obey inequality (44), if $|Sp\mathbf{Z}_{V\rho}|$ is not small. In the case of a small value of $|Sp\mathbf{Z}_{V\rho}|$, $|Sp\mathbf{Z}_{V\rho}| \to 0$, the $\kappa_1$-value is given by Eq. (43); this case corresponds to the pressure $P_0 \approx -(2/3)P_L^*$.

### 4.3. Fast heat- and mass exchanges; $(V,u)$-theory

Further the case of large values of the parameters $\lambda_{TT}$ and $\lambda_{\rho\rho}$, $\lambda_{TT}, \lambda_{\rho\rho} \gg z_{uu}, |z_{VV}|$, is considered. In the $(V, \rho, T)$-theory, this case corresponds to the one-variable $(V)$-theory. Dividing Eq. (42) by $\lambda_{\rho\rho}$ and passing to the limit $\lambda_{\rho\rho} \to \infty$, we get the following characteristic equation:

$$\kappa^2 - z_{uu}\kappa + z_{uV} = 0 \tag{47}$$

from where

$$\kappa_1 = \frac{1}{2}\left\{z_{uu} - \sqrt{z_{uu}^2 + 4|z_{uV}|}\right\} \tag{48}$$

Eq. (47) is the characteristic equation of the two-variable $(V, u)$-theory which is a generalization of the one-variable $(V)$-theory to the case of arbitrary viscosity. Accordingly, Eq. (48) involves the limiting cases of high and low viscosity.

In the case of high viscosity, $z_{uu}^2 \gg |z_{uV}|$, or, equivalently, $z_{uu} \gg |z_{VV}|$, Eq. (48) gives

$$\kappa_1 = z_{VV} \tag{49}$$

which is the result of the one-variable $(V)$-theory [35]. In the case of low viscosity, $z_{uu}^2 \ll |z_{uV}|$, $z_{uu} \ll |z_{VV}|$, Eq. (48) transforms into Eq. (46), or Eq. (B13).

Transform the preexponential factor in Eq. (36) for the nucleation rate in the case of low viscosity, Eq. (B13). Substituting $|\kappa_1| = \sqrt{|z_{uV}|}$ into Eq. (36), we obtain

$$I = \sqrt{\frac{kT_0}{2\pi M}} f_0^* = \bar{u} f_0^*, \quad f_0^* = N_b e^{-W_*/kT_0} \tag{50}$$

where the bubble "mass" $M$ is given by Eq. (19), $f_0^*$ is the density of critical bubbles, $\bar{u} = \sqrt{kT_0/2\pi M}$ is the mean thermal velocity of a particle with the mass $M$ moving towards the barrier. Thus, we have the case of Eyring's theory [43] (the transition state method) in which the inertial motion of a particle without damping leads to the independence of the flux over the barrier of the form of the barrier top. Also the nucleation rate, Eq. (50), *does not depend on viscosity*, whereas in the one-dimensional case of high viscosity, Eq. (49), the nucleation rate is inversely proportional to it. It is interesting that the nucleation rate, Eq. (50), is determined by thermodynamics only and does not depend on kinetic parameters of the liquid at all. The "velocity" $\bar{u}$ plays the role of kinetic factor. The normalizing factor $N_b$ is determined in the framework of statistical-mechanical approach [27-30].

The one-dimensional nucleation-rate formula corresponding to high viscosity is in common use in practice. As follows from the foregoing, it qualitatively differs from the formula for low viscosity, Eq. (50). Quantitatively, it gives overestimated values of the nucleation rate for liquids

1313with moderate and especially low viscosity. For the ratio of nucleation rates in the limiting cases of low and high viscosity, we have the following equation, in view of Eqs. (36), (46) and (49):

$$\frac{I_{z_{uu} \to 0}}{I_{z_{uu} \to \infty}} = \sqrt{\frac{z_{uu}}{|z_{VV}|}} = 4\eta_0 \sqrt{\frac{1}{2\sigma\rho_0 R_*}} \tag{51}$$

As an example, the estimates of the parameter $\varsigma = z_{uu}/|z_{VV}| = 8\eta_0^2/\sigma\rho_0 R_*$ for water can be made: $\eta_0 = 3\times 10^{-3}\ g\ cm^{-1}s^{-1}$ for $T_0 = 100\ °C$, $\eta_0 = 10^{-2}\ g\ cm^{-1}s^{-1}$ for $T_0 = 20\ °C$; $\sigma = 70\ erg/cm^2$, $\rho_0 = 1\ g/cm^3$. First of all, the molecular volume $\upsilon_0$ and the mean intermolecular distance $a$ are estimated as follows: $\upsilon_0 = 18/N_A \approx 3\times 10^{-23}\ cm^3$, where $N_A$ is Avogadro's number, $a = \sqrt[3]{\upsilon_0} \approx 3\times 10^{-8}\ cm$. The mentioned parameter is

$$\varsigma \approx \frac{10^{-6}}{R_*}\ \text{for}\ T_0 = 100\ °C,\ \varsigma \approx \frac{10^{-5}}{R_*}\ \text{for}\ T_0 = 20\ °C \tag{52}$$

It is seen that the limit of high viscosity, $\varsigma \gg 1$, for water at $T_0 = 100\ °C$ can be used only in the region of high supersaturation, $R_* \sim a$. For larger values of $R_*$, this criterion is not satisfied. Of course, these estimates do not take into account the dependence of the surface tension on radius. In Ref. [41], the following dependence $\sigma(R)$ at small $R$ has been obtained:

$$\sigma = KR \tag{53}$$

where $K$ is the coefficient depending on temperature. The decrease in the surface tension at small bubble size increases $\varsigma$.

In Ref. [35], $\ddot{V} = 0$ was put in Eq. (15) in the case of high viscosity and the following equation for $\dot{V}$ was used:

$$\dot{V} = 3\xi V\left(P - \frac{2\sigma}{\sqrt{b}V^{1/3}} - P_0\right) \tag{54}$$

In order to derive the criterion for neglecting the term $\ddot{V}$, we present it in the form

$$\ddot{V} = \dot{V}\frac{d\dot{V}}{dV} \tag{55}$$

Neglecting by the term $\sim \dot{V}^2$, as before, in Eq. (15), we obtain

$$\dot{V} = \left\{1 + \frac{b\rho_0 V^{2/3}}{4\eta_0}\frac{d\dot{V}}{dV}\right\}^{-1} 3\xi V\left(P - \frac{2\sigma}{\sqrt{b}V^{1/3}} - P_0\right) \tag{56}$$

So, the desired criterion is

$$\frac{b\rho_0 V_*^{2/3}}{4\eta_0}\left|\frac{d\dot{V}}{dV}\right|_* \ll 1 \tag{57}$$

Substituting here $\left|d\dot{V}/dV\right|_* = \left|z_{VV}\right| = \xi P_L^* = \sigma/2\eta_0 R_*$, one obtains finally

$$\frac{8\eta_0^2}{\sigma\rho_0 R_*} = \frac{z_{uu}}{|z_{VV}|} = \varsigma \gg 1 \tag{58}$$

which is just the criterion of high viscosity.

**4.4. On applicability of the theory to cavitation in helium II**

Helium-4 above the $\lambda$-point (helium I) is a usual liquid with respect to boiling, so the presented theory can be directly applied to it. Some notes should be made for helium-4 below the $\lambda$-point (helium II). In view of high heat conductivity of superfluid helium-4, the limiting case $\lambda_{TT} \to \infty$ considered above and resulting in Eq. (43) is just relevant for this liquid. Also for this reason, the local overheating (the temperature fluctuation) as a center of the bubble nucleation [8] is impossible. Another source of the bubble is the local rarefaction (the density fluctuation) which is stimulated by a negative pressure; just the nucleation of bubbles under negative pressures is observed and studied in helium II [11-17]. However, in order to apply the presented theory to helium II, some points relating to the dynamics of a spherical cavity should be clarified. As is known, this liquid consists of two components – normal ($n$) and superfluid ($s$); hydrodynamics of helium II has been developed by Landau in Refs. [44, 45]. The problem of motion of incompressible helium II was shown in Ref. [45] to reduce to the solution of two problems of the usual hydrodynamics – for perfect and viscous fluids. In other words, the superfluid component moves according to Euler's equation

$$\frac{\partial \mathbf{v}_s}{\partial t} + (\mathbf{v}_s \nabla)\mathbf{v}_s = -\frac{1}{\rho_s}\nabla P_s \tag{59}$$

whereas the Navier-Stokes equation holds for the normal component:

$$\frac{\partial \mathbf{v}_n}{\partial t} + (\mathbf{v}_n \nabla)\mathbf{v}_n = -\frac{1}{\rho_n}\nabla P_n + \frac{\eta_n}{\rho_n}\Delta \mathbf{v}_n \tag{60}$$

Here $\rho_s$, $\mathbf{v}_s$ and $\rho_n$, $\mathbf{v}_n$ are the mass densities and velocities of the superfluid and normal components, $\rho_n + \rho_s = \rho_0$; $\eta_n$ is the viscosity of the normal motion[46]; the pressure in helium II is presented in Ref. [45] as the sum of "the pressures of superfluid and normal motions", $P_s$ and $P_n$, respectively.

The continuity equations are

$$\text{div }\mathbf{v}_s = 0, \quad \text{div }\mathbf{v}_n = 0 \tag{61}$$

For a spherical cavity of radius $R$, the solution of Eqs. (61) subject to the boundary condition $v(R) = dR/dt$ has the form

$$v_s(r) = v_n(r) \equiv v(r) = \left(\frac{R}{r}\right)^2 \frac{dR}{dt} \tag{62}$$

Substituting this solution into Eqs. (59), (60) and integrating them over $r$ from $R$ to $\infty$, one obtains

$$\rho_s\left[R\ddot{R} + \frac{3}{2}\dot{R}^2\right] = P_{s,R} - P_{s,\infty} \tag{63a}$$

$$\rho_n\left[R\ddot{R} + \frac{3}{2}\dot{R}^2\right] = P_{n,R} - P_{n,\infty} \tag{63b}$$

where $P_{s,R}$ and $P_{s,\infty}$ are the pressures of the superfluid component at the cavity boundary and far from the cavity; the same is true for the normal component pressures $P_{n,R}$ and $P_{n,\infty}$. Obviously, $P_{s,\infty} + P_{n,\infty} = P_0$. The boundary condition to Eqs. (63a) and (63b) is the equality of the normal stresses at the cavity boundary:

$$\left(P_{n,R} - 2\eta_0 \frac{\partial v_n(r)}{\partial r}\bigg|_{r=R}\right) + P_{s,R} = P - P_L \tag{64a}$$

with

$$\eta_0 = \frac{\rho_n}{\rho_0}\eta_n \tag{64b}$$

where $P$ is the vapor pressure, as before, and $P_L = 2\sigma/R$ is the Laplace pressure. The pressure $P_n$ sufficiently far from the transition temperature $T_\lambda$ is the pressure of the ideal gas of excitations (phonons and rotons) [44], $P_n \sim \rho_n kT = (\rho_n/\rho_0)\rho_0 kT$. The fraction $\rho_n/\rho_0$ of the normal component depends on temperature and tends to zero at $T \to 0$ [44]. The "effective viscosity" $\eta_0$ in the form of Eq. (64b) provides the simultaneous disappearance of both the addends in brackets of Eq. (64a) at $T \to 0$, when the normal component vanishes.

Adding Eqs. (63a) and (63b) and employing Eq. (64a), we get

$$R\ddot{R} + \frac{3}{2}\dot{R}^2 + \frac{4\eta_0}{\rho_0}\frac{\dot{R}}{R} = \frac{1}{\rho_0}[P - P_L - P_0] \tag{65}$$

This equation has the same form as for a cavity in a viscous liquid, Eq. (64) of Ref. [35], with constant density $\rho_0$ and viscosity $\eta_0$ strongly depending on temperature, according to Eq. (64b). At $T \to T_\lambda$, $\rho_n/\rho_0 \to 1$ and $\eta_0 \to \eta_n$, so Eq. (65) describes the cavity dynamics in helium I.

The estimates for helium II similar to those for water can be made also. The following parameters from literature are used: $P_0 = -8\,bar = -8\times 10^6\,dyn/cm^2$, $T = 1.1\,°K$, $\rho_n/\rho_0 \approx 0.01$, $\sigma = 0.35\,erg/cm^2$, $\eta_n = 2\times 10^{-5}\,g\,cm^{-1}\,s^{-1}$, $\rho_0 = 0.145\,g/cm^3$, $m = 6.64\times 10^{-24}\,g$ is the atomic mass. The limit of high negative pressure, $|P_0| \gg P_*$, holds, so the critical radius is determined by the equation $R_* = 2\sigma/|P_0| \approx 10^{-7}\,cm \approx 3a$, where $a \approx 3.6\times 10^{-8}\,cm$ is the mean interatomic spacing. The parameter $\lambda_{\rho\rho}$ is determined in Ref. 35: $\lambda_{\rho\rho} = (3/4)\beta u(T_0)/R_*$, where $u(T_0) = \sqrt{8kT_0/\pi m} \approx 7\times 10^3\,cm/s$ and $\beta$ is the condensation coefficient; the value of $\beta = 0.9 \div 1$ for helium II has been reported in Ref. [47]. The parameters $z_{VV}$, $z_{uu}$ and $\lambda_{\rho\rho}$ are estimated as follows: $|z_{VV}| = \sigma/2\eta_0 R_* \approx 10^{13}\,s^{-1}$, $z_{uu} = 4\eta_0/\rho_0 R_*^2 \approx 5.5\times 10^8\,s^{-1}$, $\lambda_{\rho\rho} \approx 5\times 10^{10}\,s^{-1}$.

So, we have a low-viscosity case, $z_{uu} \ll |z_{VV}|$, for the given temperature. However, the parameter $\lambda_{\rho\rho}$ is large in comparison with $z_{uu}$, so Eq. (43) must be employed for calculating the nucleation rate. The only simplification in this equation, in view of the inequalities $z_{uu} \ll \lambda_{\rho\rho} \ll z_{VV}$ and the condition of high negative pressure, $P_L^* \sim P_0 \gg P_*$, is $Sp\mathbf{Z}_{V\rho} = z_{VV}$ and the square root has the form $\sqrt{\lambda_{\rho\rho}^2 + 3|z_{uV}|}$; as follows from the above estimates, $\lambda_{\rho\rho}^2 \sim |z_{uV}|$. If we assume the same values of the parameters for helium I ($T > T_\lambda$, $\rho_n/\rho_0 = 1$), then one obtains $|z_{VV}| \approx 10^{11}\,s^{-1}$ and $z_{uu} \approx 5.5\times 10^{10}\,s^{-1}$, $|z_{VV}| \sim z_{uu}$, i.e. an intermediate case holds.

It should be recalled that the condition of high negative pressure (which is equivalent to the case of a non-volatile liquid, $P_* \to 0$) in the high-viscosity ($V,\rho$)-theory [35] leads to the one-dimensional theory with $\kappa_1 = z_{VV}$. We see that Eq. (43) of the ($V,\rho,u$)-theory under the same condition contains all the kinetic parameters (except $\lambda_{TT}$), $z_{VV}$, $z_{uu}$ and $\lambda_{\rho\rho}$, i.e. the theory does not reduce to a one-dimensional form. In other words, the nucleation theory for a liquid with arbitrary viscosity remains multivariable even for bubbles with small critical radii $R_*$. However, the question of the dependence $P_*(R_*)$ for small bubbles (see the remark after Eq. (8)) does not arise, simply because the quantity $P_*$ is neglected. Thus, the case of arbitrary viscosity is fundamentally different from the case of high viscosity.

## 5. Conclusions

The presented theory complements the previous high-viscosity $\{V,\rho,T\}$-theory extending it to arbitrary values of viscosity. Thus, it covers all possible experimental situations with arbitrary relations between the kinetic parameters controlling the nucleation process: viscosity, inertia of a liquid, the rate of evaporation into a bubble and the rate of heat exchange between the bubble and ambient liquid. Differently from the $\{V,\rho,T\}$-theory, the arbitrary-viscosity theory does not reduce to a one-dimensional form in the case of high negative pressure which corresponds to small critical radii. This fact shows the importance of the presented multivariable theory for calculating the nucleation rates in real liquids.

The kinetic limits corresponding to specific relations between the mentioned parameters have been considered. As well as in the $\{V,\rho,T\}$-theory, they provide insight to the kinetics of homogeneous nucleation and lead into better understanding of the model. The limit of high heat exchange with Eq. (43) is relevant for helium II. The same limit together with low viscosity and low mass exchange parameter results in conclusion that nucleation can occur only under negative pressures. In the limit of low viscosity (under the conditions of Appendix B), the nucleation rate does not depend on kinetic parameters (including the viscosity itself) and is determined by thermodynamics only.

## Appendix A: high-viscosity limit

Assume that the quantities $|z_{VV}|$, $\lambda_{\rho\rho}$, $\lambda_{TT}$ are of the same order of magnitude and $z_{uu} \gg |z_{VV}|, \lambda_{\rho\rho}, \lambda_{TT}$. We introduce the small parameters

$$\beta_1 = \frac{Sp\mathbf{Z}_{(3)}}{z_{uu}}, \quad \beta_2 = \frac{\det \mathbf{\Lambda}}{z_{uu}^2}, \quad \beta_2' = \frac{B}{z_{uu}^2}, \quad \beta_3 = \frac{\det \mathbf{Z}_{(3)}}{z_{uu}^3} \tag{A1}$$

$\beta_1 \ll 1$, $\beta_2 \sim \beta_2' \sim \beta_1^2$, $\beta_3 \sim \beta_1^3$.

We have for the quantities $p$, $q$, and $r$, Eqs. (34a)-(34c), in view of Eqs. (25a)-(25d),

$$p = -\frac{3}{8}z_{uu}^2\left\{1 - \frac{8}{3}\beta_1 - \frac{8}{3}\beta_2\right\}$$

$$q = -\frac{1}{8}z_{uu}^3\left\{1 - 4\beta_1 - 4\beta_2 + 8\beta_2'\right\}$$

$$r = -\frac{3}{256} z_{uu}^4 \left\{ 1 - \frac{16}{3}\beta_1 - \frac{16}{3}\beta_2 + \frac{64}{3}\beta_2' - \frac{256}{3}\beta_3 \right\} \tag{A2}$$

In the Descartes-Euler solution, the combination of signs in the expression $\pm\sqrt{y_1} \pm \sqrt{y_2} \pm \sqrt{y_3}$ is chosen to satisfy the condition [42]

$$\left(\pm\sqrt{y_1}\right)\left(\pm\sqrt{y_2}\right)\left(\pm\sqrt{y_3}\right) = -\frac{q}{8} \tag{A3}$$

From Eq. (A2), $q < 0$, hence the product in Eq. (A3) is positive: three roots contain two minuses and one plus in various combinations, the fourth root contains three pluses. It is not difficult to prove the following inequalities for the quantities $y_1$, $y_2$, and $y_3$, Eqs. (30a)-(30c):

$$y_1 > y_2 > y_3 \tag{A4}$$

It follows from these considerations that the single negative root of Eq. (23) is given by Eq. (29).

The next step is to find the quantities $p'$ and $q'$, Eqs. (35a), (35b), with the use of Eq. (A2). Up to quadratic terms, $p'$ has the form

$$p' = \frac{z_{uu}^4}{16}\left\{\beta_2' - \frac{1}{3}\beta_1^2\right\} = \frac{z_{uu}^4}{16} Y, \quad Y = B - \frac{1}{3}\left(Sp\mathbf{Z}_{(3)}\right)^2 \tag{A5}$$

The quantity $q'$ up to cubic terms is

$$q' = \frac{z_{uu}^6}{64}\left\{-\frac{2}{27}\beta_1^3 + \frac{1}{3}\beta_1\beta_2' - \beta_3\right\} = \frac{z_{uu}^6}{64} U, \quad U = -\frac{2}{27}\left(Sp\mathbf{Z}_{(3)}\right)^3 + \frac{1}{3}BSp\mathbf{Z}_{(3)} - \det\mathbf{Z}_{(3)} \tag{A6}$$

The same designations, $Y$ and $U$, are employed in Ref. [35] for the solution of cubic equation of the $(V,\rho,T)$-theory, Eq. (28).

With the use of Eqs. (A5) and (A6), $\cos\phi'$, Eq. (31), can be determined:

$$\cos\phi' = -\frac{U}{2\sqrt{-(Y/3)^3}} = \cos\phi \tag{A7}$$

where $\cos\phi$ determines the solution of the characteristic equation in the $(V,\rho,T)$-theory. Thus, $\cos\phi' = \cos\phi$.

The quantities $y_1$, $y_2$, and $y_3$, Eqs. (30a)-(30c), up to linear terms are

$$y_1 = \frac{z_{uu}^2}{16}\left\{1 - \frac{8}{3}\beta_1 + \frac{8A_1}{z_{uu}}\right\}$$

$$y_2 = \frac{z_{uu}^2}{16}\left\{1 - \frac{8}{3}\beta_1 - \frac{8A_2}{z_{uu}}\right\}$$

$$y_3 = \frac{z_{uu}^2}{16}\left\{1 - \frac{8}{3}\beta_1 - \frac{8A_3}{z_{uu}}\right\} \tag{A8}$$

where

$$A_1 \equiv \sqrt{-\frac{Y}{3}}\cos\frac{\phi}{3}, \quad A_2 \equiv \sqrt{-\frac{Y}{3}}\cos\left(\frac{\phi}{3}+\frac{\pi}{3}\right), \quad A_3 \equiv \sqrt{-\frac{Y}{3}}\cos\left(\frac{\phi}{3}-\frac{\pi}{3}\right)$$

Substituting Eq. (A8) into Eq. (29) for $\kappa_1$ and expanding the square roots up to linear terms, we find

$$\kappa_1 = \frac{z_{uu}}{3}\beta_1 + A_2 - A_1 - A_3 \quad (A9)$$

In view of the identity

$$A_2 - A_1 - A_3 = -2\sqrt{-\frac{Y}{3}}\cos\left(\frac{\phi}{3}-\frac{\pi}{3}\right) \quad (A10)$$

one obtains finally

$$\kappa_1 = \frac{1}{3}\left\{-2\sqrt{(Sp\mathbf{Z}_{(3)})^2 - 3B}\cos\left(\frac{\phi}{3}-\frac{\pi}{3}\right) + Sp\mathbf{Z}_{(3)}\right\} \quad (A11)$$

i.e. the negative eigenvalue of the $(V,\rho,T)$-theory [35] (the solution of Eq. (28)), as it must; $\cos\phi$ is determined by Eq. (A7) with Eqs. (A5) and (A6) for $Y$ and $U$. This result confirms the correctness of Eq. (29) as well as the self-consistency of the theory.

## Appendix B: low-viscosity limit

As before, the quantities $|z_{VV}|$, $\lambda_{\rho\rho}$, $\lambda_{TT}$ are assumed to be of the same order of magnitude and $z_{uu} \ll |z_{VV}|, \lambda_{\rho\rho}, \lambda_{TT}$. Represent the quantities $p$, $q$, and $r$, Eqs. (34a)-(34c), in the form of an expansion in some small parameters. Employing Eqs. (25a)-(25d), it is not difficult to get the following expressions up to linear terms:

$$p = p_0(1+p_1), \quad p_0 = -\frac{3}{8}(\lambda_{\rho\rho}+\lambda_{TT})^2 + \det\mathbf{\Lambda}, \quad p_1 = \frac{1}{p_0}\left\{Sp\mathbf{Z}_{(3)} - \frac{3}{4}(\lambda_{\rho\rho}+\lambda_{TT})\right\}z_{uu} \quad (B1)$$

$$q = q_0(1+q_1), \quad q_0 = -\frac{1}{8}(\lambda_{\rho\rho}+\lambda_{TT})^3 + \frac{1}{2}(\lambda_{\rho\rho}+\lambda_{TT})\det\mathbf{\Lambda},$$

$$q_1 = \frac{1}{q_0}\left\{-\frac{3}{8}(\lambda_{\rho\rho}+\lambda_{TT})^2 + \frac{1}{2}\left[(\lambda_{\rho\rho}+\lambda_{TT})Sp\mathbf{Z}_{(3)} + \det\mathbf{\Lambda}\right] - B\right\}z_{uu} \quad (B2)$$

$$r = r_0(1+r_1), \quad r_0 = -\frac{3}{256}(\lambda_{\rho\rho}+\lambda_{TT})^4 + \frac{1}{16}(\lambda_{\rho\rho}+\lambda_{TT})^2\det\mathbf{\Lambda},$$

$$r_1 = \frac{1}{r_0}\left\{-\frac{3}{64}(\lambda_{\rho\rho}+\lambda_{TT})^3 + \frac{1}{16}\left[(\lambda_{\rho\rho}+\lambda_{TT})^2 Sp\mathbf{Z}_{(3)} + 2(\lambda_{\rho\rho}+\lambda_{TT})\det\mathbf{\Lambda}\right]\right\}$$

$$-\frac{1}{4}(\lambda_{\rho\rho}+\lambda_{TT})B+\det \mathbf{Z}_{(3)}\bigg\}z_{uu} \tag{B3}$$

It is seen that $p_1 \sim q_1 \sim r_1 \ll 1$.

The quantities $p'$ and $q'$, Eqs. (35a), (35b), in view of Eqs. (B1)-(B3), can be represented similarly:

$$p' = -\frac{1}{4}p'_0(1+p'_1), \quad p'_0 = \frac{p_0^2}{12}+r_0, \quad p'_1 = \frac{1}{p'_0}\left\{\frac{p_0^2}{6}p_1 + r_0 r_1\right\} \tag{B4}$$

$$q' = -\frac{1}{8}q'_0(1+q'_1), \quad q'_0 = \frac{1}{108}p_0^3 - \frac{1}{3}p_0 r_0 + \frac{1}{8}q_0^2, \quad q'_1 = \frac{1}{q'_0}\left\{\frac{3p_0^3}{108}p_1 - \frac{p_0 r_0}{3}(p_1+r_1)+\frac{q_0^2}{4}q_1\right\} \tag{B5}$$

The quantities $p'_0$ and $q'_0$ are easily calculated with the use of Eqs. (B1)-(B3):

$$p'_0 = \frac{1}{12}(\det \mathbf{\Lambda})^2, \quad q'_0 = \frac{1}{108}(\det \mathbf{\Lambda})^3 \tag{B6}$$

so that

$$-\frac{p'}{3} = \frac{(\det \mathbf{\Lambda})^2}{144}(1+p'_1), \quad q' = -\frac{1}{864}(\det \mathbf{\Lambda})^3(1+q'_1) \tag{B7}$$

Now $\cos\phi'$, Eq. (31), can be calculated:

$$\cos\phi' = \frac{1+q'_1}{(1+p'_1)^{3/2}} = 1+q'_1 - \frac{3}{2}p'_1 = 1 - \frac{\phi'^2}{2} \tag{B8}$$

from where

$$\phi'^2 = 3p'_1 - 2q'_1 \tag{B9}$$

After some transformations, the following expression for $\phi'^2$ is obtained:

$$\phi'^2 = \frac{108|\det \mathbf{Z}_{(3)}|}{(\det \mathbf{\Lambda})^2}\left\{\frac{(\lambda_{\rho\rho}+\lambda_{TT})^2}{4\det \mathbf{\Lambda}}-1\right\}z_{uu} \tag{B10}$$

In view of smallness of $\phi'$,

$$\cos\frac{\phi'}{3} = 1-\frac{\phi'^2}{18}, \quad \cos\left(\frac{\phi'}{3}+\frac{\pi}{3}\right) = \frac{1}{2}\left\{1-\frac{\phi'}{\sqrt{3}}-\frac{\phi'^2}{18}\right\}, \quad \cos\left(\frac{\phi'}{3}-\frac{\pi}{3}\right) = \frac{1}{2}\left\{1+\frac{\phi'}{\sqrt{3}}-\frac{\phi'^2}{18}\right\} \tag{B11}$$

Calculating $\kappa_1$, Eq. (29), in a way similar to Appendix A, we find

$$-\sqrt{y_1} = -\frac{\lambda_{\rho\rho}+\lambda_{TT}}{4}+O(z_{uu}), \quad -\sqrt{y_2}+\sqrt{y_3} = -\frac{\det \mathbf{\Lambda}}{3\sqrt{(\lambda_{\rho\rho}+\lambda_{TT})^2-4\det \mathbf{\Lambda}}}\frac{\phi'}{\sqrt{3}}+O(z_{uu}) \tag{B12}$$

Substituting Eq. (B12) into Eq. (29) and employing Eq. (B10) and the equality $\det \mathbf{Z}_{(3)} = z_{VV}\det \mathbf{\Lambda}$, we get finally

$$\kappa_1 = -\sqrt{|z_{VV}|z_{uu}}+O(z_{uu}) = -\sqrt{|z_{uV}|} \tag{B13}$$